\documentstyle[12pt,epsf]{article}
\begin{document}
\begin{titlepage}
\begin{flushright}
KEK-CP-088 \\
\end{flushright}
\vskip 2cm
\begin{center}
{\Large
QED Radiative Corrections to the Non-annihilation Processes
Using the Structure Function and the Parton Shower
}
\end{center}
\vskip 1cm
\begin{center}
{\large Y. Kurihara, J. Fujimoto, Y. Shimizu} \\
{\it High Energy Accelerator Research Organization,\\
Oho 1-1, Tsukuba, Ibaraki 305-0801, Japan}\\
\vskip 5mm
{\large K. Kato, K. Tobimatsu} \\
{\it Kogakuin University, Shinjuku, Tokyo 163-8677, Japan} \\
\vskip 5mm
{\large T. Munehisa} \\
{\it Yamanashi University, Kofu, Yamanashi 400-8510, Japan}
\end{center}
\vskip 1cm
\begin{abstract}
Inclusion of the QED higher order radiative corrections in the two-photon 
process, $e^+e^- \rightarrow e^+e^-\mu^+\mu^-$, is examined by means of
the structure function and the parton shower. Results are compared with 
the exact $O(\alpha)$ calculations and give a good agreement. These two 
methods should be universally applicable to any other non-annihilation 
processes like the single-$W$ productions in the $e^+e^-$ collisions. 
In this case, however, the energy scale for the evolution by 
the renormalization-group equation should be chosen properly depending on 
the dominant diagrams for the given process. A method to find the most 
suitable energy scale is proposed. 
\end{abstract}
\end{titlepage}

\section{Introduction}
A precise prediction of the cross sections for high-energy $e^+e^-$ 
scattering frequently requires an estimation of the radiative 
corrections beyond the lowest order calculations. Among various 
corrections from the electro-weak interactions it is known that 
the QED radiative corrections to the initial-state particles gives 
the greatest contribution in general. Sometimes all order summation 
is necessary to 
give the needed precision. In the leading-logarithmic approximation
the higher-order summation of the QED corrections can be done easily 
thanks to the factorization theorem. The theorem given by ref.\cite{kln} 
guarantees that the final state corrections result only a small 
contribution to the total cross section.

For the $e^+ e^-$ annihilation processes we have tools. They are
universally applicable because only the initial-state QED corrections 
are involved; the structure function\cite{SF} and 
the parton shower\cite{PS,QEDPSs} methods widely used in high energy 
physics today. The recent experiments, however, require the higher order 
corrections for the multi-particle final states, such as the four-fermion 
productions at LEP2. Even though the exact calculations of the higher 
order corrections are very difficult or impossible, still it is possible
to include the biggest QED corrections by making use of those 
tools as long as the process is dominated by the annihilation.

On the other hand for the non-annihilation processes it has been not well
investigated how to apply these tools. Only a few examples are the Bhabha 
scattering to which the structure function has been used  
in\cite{bbm} and the parton shower in\cite{QEDPSt}. 
In the present work it will be shown that
the structure function and the parton shower can be also the universal 
tools for any non-annihilation process once the evolution energy is 
settled. 

First we apply and examine both methods to the two-photon process, 
$e^+e^- \rightarrow e^+e^-\mu^+\mu^-$ in the next section. The obtained
total and differential cross sections are compared with those given 
by the $O(\alpha)$ corrections\cite{bdk} in section 3.
These methods must be universal to any process, again thanks to 
the factorization theorem. The energy scale, however, with which 
the radiative correction is evolved should be carefully chosen 
for the non-annihilation processes. An unique and definite way to 
find this energy scale is explained in section 4.

\section{Calculation method}
\subsection{Structure Function Method}
The structure function(SF) for the initial state radiative correction(ISR)
is well established for the case of the $e^+e^-$ {\sl annihilation} 
processes. The observed cross section corrected by ISR can be expressed 
by using SF as
\begin{eqnarray}
\sigma_{total}(s)&=&\int^{1} dx_1 \int^{1} dx_2 D_{e^-}(x_1,s)D_{e^+}(x_2,s)
\sigma_0(x_1 x_2 s),
\label{SFs}
\end{eqnarray}
where $D_{e^\pm}(x,s)$ is the electron(positron) structure function 
with $x_{1,2}$ being the energy fractions of $e^\pm$. The SF up to the
$O(\alpha ^2)$ is given by\cite{jf}
\begin{eqnarray}
D(x,s) &=&\frac{\beta}{2}\left(1-x\right)^{\frac{\beta}{2}-1} \nonumber
\left[1+\frac{3}{8}\beta +\beta^2\left(\frac{9}{128}
-\frac{\pi^2}{48}\right)\right] 
-\frac{\beta}{4}\left(1+x\right) \\ 
&+&\left(\frac{\beta}{4} \right)^2\left[2(1+x){\tt ln}\frac{1}{1-x} 
- \frac{1+3x^2}{2(1-x)} {\tt ln}x -\frac{5+x}{2}\right], \\ 
\beta&=&\frac{2 \alpha}{\pi}\left({\tt ln}\frac{s}{m_e^2}-1\right).
\end{eqnarray}
In deriving this formula, which is given by solving the Altarelli-Parisi 
equation in the LL approximation\cite{AP}, we have used an {\it ad hoc} 
trick to get more accuracy. 
The factor $\beta=(2 \alpha/\pi) {\tt ln}(s/m_e^2)$ is replaced by 
$\beta=(2 \alpha/\pi)({\tt ln}(s/m_e^2)-1)$ to match with 
the perturbative calculations.

Let us apply the SF method to the two-photon process,
\begin{eqnarray}
e^-(p_-)+e^+(p_+) &\rightarrow& e^-(q_-)+e^+(q_+)+\mu^-(k_-)+\mu^+(k_+).
\label{pc1}
\end{eqnarray}
For the forward scattering of $e^\pm$, the multi-peripheral diagrams 
shown in Fig.1 give the dominant contribution to the total cross section. 
Thus only the multi-peripheral diagrams are taken into account in this 
work. In this case the corrected cross section is given by
\begin{eqnarray}
\sigma_{total}(s,t_\pm) &=&\int dx_{I-}\int dx_{F-}
\int dx_{I+} \int dx_{F+}\nonumber 
D_{e^-}(x_{I-},Q_-^2) D_{e^-}(x_{F-},Q_-^2) \\
&~&D_{e^+}(x_{I+},Q_+^2) D_{e^+}(x_{F+},Q_+^2) 
\sigma_0({\hat s},{\hat t_\pm}). \label{eq1}
\end{eqnarray}
Here $Q_\pm^2$ is the energy scale to be fixed, with which SF should 
be driven. Since these functions are common to the initial and 
the final radiations from the $e^\pm$'s 
in the leading-logarithmic(LL) approximation, we shall drop the 
subscript $e^\pm$ from the SF hereafter. After(before) the photon 
radiation the initial(final) momenta $p_\pm$ ($q_\pm$) become
${\hat p_\pm}$ (${\hat q_\pm}$) in the following ways
\begin{eqnarray}
{\hat p_-} &=& x_{I-}p_-,~~~{\hat q_-}=\frac{1}{x_{F-}}q_-, \\
{\hat p_+} &=& x_{I+}p_+,~~~{\hat q_+}=\frac{1}{x_{F+}}q_+, 
\end{eqnarray}
respectively. Then the CM energy squared($s=(p_- + p_+)^2$) and 
the momentum transfer squared($t_\pm=(p_\pm - q_\pm)^2$) are scaled as
follows,
\begin{eqnarray}
{\hat s} = x_{I-}x_{I+}s,~~~{\hat t_\pm} &=& \frac{x_{I\pm}}{x_{F\pm}}t_\pm.
\end{eqnarray}
Note that SF behaves like $\delta(1-x)$ when $\beta \rightarrow
0$, that is,
\begin{eqnarray}
\int^1_0 dx \frac{\beta}{2} (1-x)^{\frac{\beta}{2}-1} f(x) &=&
f(1)+\int^1_0 dx{\biggl [}(1-x)^{\frac{\beta}{2}}-1{\biggr ]}f'(x) \nonumber \\
&=& f(1) + \frac{\beta}{2} \int^1_0 dx {\tt ln}(1-x)f'(x)+O(\beta^2),
\end{eqnarray}
where $f(x)$ is an arbitrary smooth function.

The choice of the energy scale in SF is not a trivial matter. 
As pointed out in Ref.\cite{bbn} it is natural to use $-t_\pm$ instead 
of $s$ for the two-photon processes. The justification of this choice 
is given by comparing them with the perturbative calculations. This
could be done in the region where the soft-photon approximation of 
the total cross section is valid. Since the corrections for the $e^-$ 
and $e^+$ sides must be symmetric, it is enough to consider only those 
from the $e^-$ side. The total correction will be obtained by doubling 
them. Then Eq.(\ref{eq1}) can be simplified as
\begin{eqnarray}
\sigma(s)&=&\int dx_I \int dx_F D(x_I,Q^2) D(x_F,Q^2) \sigma_0(x_I s).
\end{eqnarray}
The integrations are performed in the region $1-x_{I,F} \ll 1$.
Since the Born cross section of this process($\sigma_0({\hat s})$) is 
a smooth function of ${\hat s}$, the cross section in the soft photon 
approximation is written as
\begin{eqnarray}
\sigma_{soft}&=&
\sigma_0(s)\int^1_{1-\frac{k_c}{E}} dx_I 
\int^1_{(1-\frac{k_c}{E})/x_I} dx_F  D(x_I,Q^2) D(x_F,Q^2) \nonumber \\
&=& \sigma_0(s) \int^{\frac{k_c}{E}}_0 dy H(y,Q^2), \\
H(y,Q^2) &=& \int^1_{1-y} \frac{dx_F}{x_F}
D(x_F,Q^2) D\left(\frac{1-y}{x_F},Q^2\right),
\end{eqnarray}
where $k_c$ is the maximum energy of the sum of the initial and the final
photon energies and $E=p^0 \approx q^0$. The function $H$ is called 
{\it radiator} and can be obtained easily from SF as
\begin{eqnarray}
H(x,s)&=&D(1-x,s)|_{\beta \rightarrow 2\beta} \nonumber \\
&=&\beta x^{\beta-1} \left[1+\frac{3}{4}\beta +\frac{\beta^2}{4}\left(\frac{9}{8}
-\frac{\pi^2}{3}\right) \right] - \beta\left(1-\frac{x}{2}\right) \nonumber \\ 
&+&\frac{\beta^2}{8}\left[-4(2-x){\ln}x - \frac{1+3(1-x)^2}{x}
{\ln}(1-x) -6 +x \right]. 
\end{eqnarray}
The integration of the function $H$ in the small $k_c/E$ region gives
\begin{eqnarray}
\int^{\frac{k_c}{E}}_0 dy H(y,Q^2)
&=&1+\frac{\alpha}{\pi}\left[-2l(L-1)+\frac{3}{2}(L-1)\right] +O(\alpha^2), 
\end{eqnarray}
where $L={\ln}({Q^2}/{m_e^2})$, and $l={\ln}({E}/{k_c})$.
Then the cross section in the soft photon approximation up to the
$O(\alpha)$ is obtained,
\begin{eqnarray}
\sigma_{soft} &=& \sigma_0(s)\left\{1+\frac{\alpha}{\pi}\left[-2l(L-1)
+\frac{3}{2}(L-1)\right]\right\}.
\label{eq2}
\end{eqnarray}
This expression is compared with the perturbative calculation 
given by Berends, Daverveldt and Kleiss\cite{bdk}({\tt BDK} hereafter).
In the {\tt BDK} program the multi-peripheral diagrams and their $O(\alpha)$
corrections of the self-energy correction to $e^\pm$, the vertex
correction for $e^\pm$-$e^\pm$-$\gamma$ vertex, the soft- and hard-photon
emissions and the vacuum polarization of the virtual photons 
are calculated. The corrections from the photon bridged 
between different charged lines are not given because the contributions
from the box diagrams with photon exchange between $e^+$ and $e^-$ is 
known to be small\cite{vNV}.
The LL approximation of the virtual correction factors(vertex $+$ soft
photon) is 
\begin{eqnarray}
2{\rm Re}F_1+\delta_s &\rightarrow& \frac{\alpha}{\pi} 
\left(-2l(L_t-1)+\frac{3}{2}L_t-2\right), 
\label{eq3}
\end{eqnarray}
where $L_t={\ln}({-t}/{m_e^2})$ and $t=(p_- - q_-)^2$.
By comparing Eqs.(\ref{eq2}) and (\ref{eq3}) one concludes
that the energy scale of SF should be $Q^2=t$

This comparison shows that once the proper energy scale is found
the SF can reproduce the correct evolution of the soft-photon emission 
along the electron line up to the $O(\alpha)$. However, there remains some 
mismatch with the constant term. This can be compensated by multiplying
an overall factor to the SF. This factor is usually called {\it K-factor}.
When the SF is evolved with $-t$ this factor is found to be 
$1-\alpha/2 \pi$. Then the total {\it K-factor} for both $e^+$ and $e^-$ 
must be $1-\alpha/\pi$.

It should be noted that the assumption $-t/m_e^2 \gg 1$ does not hold
for the forward scattering of the two-photon process. The LL-terms in 
the SF are no more leading if this happens. To find the region where 
the approximation is valid the LL terms of the form factor $F_1$ is 
compared with the exact form given by {\tt BDK} in Fig.2. While they 
agree well at high $Q^2$ region as expected, they show a large 
deviation in the region $Q^2/m_e^2<10$. In order to make SF well-defined, 
the energy evolution must be truncated at some point, say $L=1$.

\subsection{Parton Shower Method}
Instead of the analytic formula of the structure function, a Monte Calro 
method based on the parton shower algorithm in QED ({\tt QEDPS}) can be 
used to solve the Altarelli-Parisi equation in 
the LL approximation\cite{AP}. 
The detailed algorithm of the {\tt QEDPS} is found in Ref.\cite{QEDPSs}
for the $e^+e^-$ annihilation processes and in Ref.\cite{QEDPSt}
for the Bhabha process. For the two-photon process the energy scale 
for the parton shower evolution is also $-t$. The truncation at $L=1$ 
is again imposed. One difference between SF and {\tt QEDPS} is that 
the {\it ad hoc} replacement $L \rightarrow L-1$, which was
realized by hand for SF, cannot be done for {\tt QEDPS}. 
This causes a deviation of the {\it K-factor} 
from the SF method. The {\it K-factor} for 
{\tt QEDPS} is given by $1-4\alpha/\pi$. Another significant difference 
between these two is that {\tt QEDPS} can treat the transverse momentum 
of emitted photons correctly by imposing the exact kinematics at the
$e\rightarrow e\gamma$ splitting. It does not affect the total cross 
sections so much when the final $e^\pm$ have no cut. However, the finite 
recoiling of the final $e^\pm$ may result a large effect on the tagged 
cross sections as shown below.

In return to the the exact kinematics at 
the $e\rightarrow e\gamma$ splitting, $e^\pm$ are no more on-shell
after photon emission. On the other hand the matrix element of the hard
scattering process must be calculated with the on-shell external particles.
A trick to map the off-shell four-momenta of the initial $e^\pm$ to 
those at on-shell is needed. Following method is used in the calculations.
\begin{enumerate}
\item ${\hat s}=({\hat p}_- + {\hat p}_+)^2$ is calculated, where
${\hat p}_\pm$ are the four-momenta of the initial $e^\pm$ after the
photon emission
by ${\tt QEDPS}$. ${\hat s}$ can be positive even for the 
off-shell $e^\pm$. (When ${\hat s}$ is negative, that event is discarded.)
\item All four-momenta are generated in the rest-frame of the initial
$e^\pm$ after the photon emission. Four-momenta
of the initial $e^\pm$ in this frame are ${\tilde p}_\pm$, where
${\tilde p}_\pm=m_e^2$ (on-shell) 
and ${\hat s}=({\tilde p}_- + {\tilde p}_+)^2$
\item All four-momenta are rotated and boosted to match 
the three-momenta of ${\tilde p}_\pm$ with those of ${\hat p}_\pm$.
\end{enumerate}
This method respects the direction of the final $e^\pm$ rather than 
the CM energy of the collision. The total energy does not conserve 
because of the virtuality of the initial $e^\pm$.

\section{Numerical Calculations}
\subsection{No-cut Case}
First the total cross section of $e^+e^- \rightarrow e^+e^- \mu^+ \mu^-$ 
without any experimental cut is considered. The exact matrix element is 
generated by the {\tt GRACE} system\cite{grace}. 
Only the multi-peripheral diagrams are generated for the test of 
the approximation methods.
The phase-space integration of the matrix element squared is
carried out numerically by {\tt BASES}\cite{bases} using an adaptive
Monte Calro method. It is trivially confirmed that the Born cross 
sections agree with {\tt BDK} results within the statistical error of 
the numerical integration.

The convolution of SF or {\tt QEDPS} with the Born cross section is not 
a straightforward task because of the complicated four-body kinematics. 
For a simple two-body process such as the Bhabha scattering the momentum
transfer($-t$) can be taken as one of the integration variables 
and the Eq.(\ref{eq1}) will be easily performed then. It is, however, 
practically impossible to use this technique for the case of the
four-body kinematics. Hence we have to admit the following approximation:
\begin{enumerate}
\item Eight random numbers obtained from {\tt BASES} correspond to the
eight independent variables of the phase-space integration.
\item All the kinematical variables are determined with {\it no}
radiative effect. The $-t_\pm$'s are fixed also.
\item The radiative correction factors of Eq.(5) are determined 
from $-t_\pm$'s, $x_{I-}$ and $x_{I+}$.
\item A new CM energy is obtained by the fixed $x_{I-}$ and $x_{I+}$.
\item All the kinematical variables are re-calculated from the same set
of the random numbers and the new CM energy.
\end{enumerate}
In this method the evolution energy scale of the radiator is not exactly
the same after the re-calculation with the new CM-energy. However, this
difference must be beyond the LL order.

The total cross sections without any experimental cuts with SF 
and {\tt QEDPS} are summarized in Table~\ref{TB1}.
The {\tt BDK} program also includes the correction from the vacuum 
polarization of the photon propagator. In order to compare the SF 
and {\tt QEDPS} results with {\tt BDK}, this correction is removed.
After including the {\it K-factor} the results of our two
methods are in good agreement with the $O(\alpha)$ calculation(without 
the vacuum polarization). 
\begin{table}[htbp]
\begin{center}
\begin{tabular}{|c||c|c|c|c|} \hline
$E_{CM}$ (GeV)&$\sigma_0$ ($nb$)&$\sigma_{\tt sf}$ ($nb$)&
$\sigma_{\tt QEDPS}$ ($nb$)&$\sigma_{\tt BDK}$ ($nb$)
\\ \hline \hline
20 & 97.0(1) & 96.3(1)& 96.0(2)&  96.0(1)  \\ \hline
40 & 137.5(3)& 136.3(1)& 135.9(3)&  135.9(1)  \\ \hline
100 & 202.8(4)& 201.0(2)& 200.5(4)& 200.5(2) \\ \hline
200 & 262.0(6)& 259.8(3)& 259.1(6)& 258.8(2)  \\ \hline
\end{tabular}
\end{center}
\caption{\footnotesize Total cross sections without the vacuum 
correction. $\sigma_0$ is the tree cross section.
Last column shows the results by the {\tt BDK} program with the vertex
correction, the soft-photon correction and the hard-photon emission. 
The vacuum polarization is removed from {\tt BDK}.
The number in parenthesis shows the statistical error of the numerical
integration on the last digit.
}
\label{TB1}
\end{table}

The effect of the vacuum polarization can be included if one uses
the running QED coupling in the SF method. The $ff\gamma$ coupling 
evolved by the renormalization group equation is given by
\begin{eqnarray}
g_{ff\gamma}(-t_\pm)=g_{ff\gamma}(0)\left(1-\frac{\alpha}{3\pi}
\sum_i C_i e_i^2 \log{-t_\pm \over m_i^2}\Theta(-t_\pm - m_i^2) \right)^{-1}, 
\end{eqnarray}
where $\alpha=1/137.036$ is the QED coupling at the zero momentum
transfer, $C_i$ the color factor, $e_i$ the electric charge in unit of
the $e^+$ charge, $m_i$ the mass of the $i$-th fermion. 
The index $i$ runs over all massive fermions.
Only those fermions whose mass is greater than $-t_\pm$ are taken into 
account through the step function.
The quark masses are chosen so as to match with the vacuum polarization 
in the {\tt BDK} program. The results are shown in Table~\ref{TB2}. 
The deviation from the {\tt BDK} program is typically around 0.5\%.
\begin{table}[htbp]
\begin{center}
\begin{tabular}{|c||c|c|c|} \hline
$E_{CM}$ (GeV)&$\sigma_{\tt sf}$ ($nb$)
&$\sigma_{\tt QEDPS}$ ($nb$)&$\sigma_{\tt BDK}$ ($nb$)
\\ \hline \hline
20 & 97.5 (1)& 96.9(1)& 97.1(1)    \\ \hline
40 & 137.8 (1) & 137.1(2)& 137.3(1)  \\ \hline
100 & 203.3 (1) & 202.2(3)& 202.4(2)  \\ \hline
200 & 262.6 (1) & 261.2(3)& 261.2(2)  \\ \hline
\end{tabular}
\end{center}
\caption{\footnotesize Total cross sections with the vacuum polarization. 
The number in parenthesis shows the statistical error of the numerical
integration on the last digit.
}
\label{TB2}
\end{table}

The differential cross sections with respect to the $e^-$ energy and 
angle, the CM energy of the final four-fermions and the invariant mass 
of the $\mu^\pm$-pair at the CM energy of 200 GeV are shown in Fig.3. 
The two programs give consistent distributions.

In order to check the recoiling effect of the final $e^-$ due to 
the photon emission by {\tt QEDPS}, the $e^-$ polar angle 
is compared in Fig.4 between SF and {\tt QEDPS}. The cross sections 
with the $e^-$ angle between $10^\circ$ and $20^\circ$ are found
to be 6.00(5.80) pb for {\tt QEDPS}(SF) at the CM energy of 200 GeV. 
It is not surprising that the agreement becomes worse than
the total cross sections, because SF includes no recoiling at all.

If a wrong energy scale of $s=(p_- + p_+)^2$ is used instead of
$t_{\pm}$ as the energy evolution scale in the ISR tool, one may get
over-estimation of the ISR effect. At the CM energy of 200 GeV,
SF with the energy scale of $s$ gives the total cross section of
$257.8$nb instead of $262.6$nb with the correct energy scale.

\subsection{Single-tagging Case}
The same comparison is done for the $e^-$-tagging case. The experimental
cuts applied are: \\
For the $e^-$,
\begin{itemize}
\item $10^{\circ}<\theta_{e^-}<170^{\circ}$,
\item $E_{e^-}> 1$ GeV.
\end{itemize}
For the $\mu^\pm$,
\begin{itemize}
\item $10^{\circ}<\theta_{\mu^{\pm}}<170^{\circ}$,
\item $E_{\mu^{\pm}}> 1$ GeV,
\item $M_{\mu \mu}> 1$ GeV.
\end{itemize}
The total cross sections with the above cuts at the CM-energy of 200 GeV 
are calculated to be $1.169 \pm 0.004$pb($1.13 \pm 0.01$pb) by 
{\tt GRACE} with {\tt QEDPS}(by {\tt BDK}). The vacuum polarization, 
{\it i.e.} the running $\alpha$, is included. This small discrepancy
may come from the finite recoiling of the final $e^-$ by the soft photon 
emissions in {\tt QEDPS}. The differential cross sections are also 
compared in Fig.4. The results of {\tt GRACE} with {\tt QEDPS} are 
in good agreement with {\tt BDK}.

\section{Energy Scale Determination}
The factorization theorem for the QED radiative corrections in the LL 
approximation is valid independent of the structure of the matrix element
of the kernel process. Hence SF and {\tt QEDPS} must be applicable to 
{\it any} $e^+e^-$ scattering processes. However, the choice of the energy 
scale in SF and {\tt QEDPS} is not a trivial issue.
For a simple process like the two-photon process with only the
multi-peripheral diagrams considered so far, the evolution energy scale 
could be determined by making use of the exact perturbative calculations. 
However, this is not always possible for more complicated processes. 
Hence a way to find a suitable energy scale without knowing the exact 
loop calculations should be established somehow.

First let us look at the general consequence of the soft photon
approximation. The soft photon cross section(including both the real 
and the virtual photon effects) is given by the Born cross section 
multiplied by some correction factor in the LL order as\cite{SW}
\begin{eqnarray}
\frac{d\sigma_{soft}(s)}{d\Omega} &=& 
\frac{d\sigma_0(s)}{d\Omega} \nonumber \\
&\times& \left| {\exp}\left[-\frac{\alpha}
{\pi} {\ln}\left( \frac{E}{k_c} \right)  
\sum_{i,j} \frac{e_i e_j \eta_i \eta_j}
{\beta_{ij}} {\ln}\left(\frac{1+\beta_{ij}}{1-\beta_{ij}}\right) 
\right] \right|^2,
\label{ir} \\
\beta_{ij} &=& \left(1-\frac{m_i^2 m_j^2}
{(p_i \cdot p_j)^2}\right)^{\frac{1}{2}},
\end{eqnarray}
where $m_j$'s($p_j$'s) are the mass(momentum) of $j$-th charged particle,
$k_c$ the maximum energy of the soft photon (boundary between soft- and
hard-photons), $E$ the beam energy, and $e_j$ the electric charge 
in unit of the $e^+$ charge.
The factor $\eta_j$ is $-1$ for the initial particles and $+1$ for the
final particles. The indices ($i,j$) run over all the charged particles 
in the initial and final states. 

For the process (\ref{pc1}) one can see 
that the soft-photon factor in Eq.(\ref{ir}) with a ($p_- \cdot q_-$)-term 
reproduces Eq.(\ref{eq3}) in the LL approximation. This implies that 
one is able to read off the possible evolution energy scale in SF from 
Eq.(\ref{ir}) without explicit loop calculations. However, one may have 
a question why the energy scale $s=(p_-+p_+)^2$ does not appear in 
the soft-photon correction even they are included in Eq.(\ref{ir}).
When we applied SF to the two-photon process in the previous section, 
we have ignored those terms which come from the photon bridged 
between different charged lines. This is because the contributions
from the box diagrams with photon exchange between $e^+$ and $e^-$ is 
known to be small\cite{vNV}. Fortunately the infrared part of 
the loop correction is already included in Eq.(\ref{ir}) and no need to
know the full form of the loop diagram. 
For the two-photon processes if one looks at two terms with, for example 
($p_- \cdot p_+$)- and with ($q_- \cdot p_+$)-terms, 
the momentum of $e^-$ is almost 
the same before and after the scattering($p_- \approx q_-$). 
Only the difference appears in $\eta_j \eta_k=+1$ for a ($p_-  p_+$)-term  
and $\eta_j \eta_k=-1$ for a ($q_- p_+$)-term. Then these terms compensate 
each other after summing them up for the forward scattering which is 
the dominant kinematical region of this process. This is why the energy 
scale $s=(p_-+p_+)^2$ does not appear in the soft-photon correction 
despite the fact that it exists in Eq.(\ref{ir}).

When some experimental cuts are imposed, for example the final $e^-$ is
tagged in a large angle, this cancellation is not perfect but partial 
and the energy scale $s$ must appear in the soft-photon correction. 
In this case the annihilation type diagrams will also contribute to 
the matrix elements. Then the usual SF and {\tt QEDPS} for 
the annihilation processes are justified to be used for the ISR with 
the energy scale $s$. One can check which energy scale is dominant 
under the given experimental cuts by numerically integrating 
the soft-photon cross section given by Eq.(\ref{ir}) over the allowed
kinematical region. Thus in order to determine the energy scale it is 
sufficient to know the infrared behavior of the radiative process using 
the soft-photon factor. In some region of the phase-space two or more 
energy scales may be involved in the soft-photon cross section with
comparable amount of contribution. In this region a simple SF and 
{\tt QEDPS} are not applicable.

\section{Conclusions}
Two practical tools to incorporate with the QED radiative corrections 
are developed for non-annihilation processes by means of the structure 
function and the parton shower. These programs are applied to the 
two-photon process, $e^+e^- \rightarrow e^+e^- \mu^+ \mu^-$. The results 
are compared with the perturbative calculation of the $O(\alpha)$ 
and show a good agreement. 

These tools should be applicable to any non-annihilation process 
universally. It is demonstrated that the energy scale for the evolution, 
which depends on the dominant diagrams in the interested kinematical 
region, can be determined with the help of the well known formula of 
the soft photon factor. As an example we have tried a real $W$ production
in $e^+e^-$ annihilation. The application of these tools to more 
complicated processes like four-fermion final state including 
a single-$W$ production are left to the future publications.


\newpage
\begin{figure}[htb]
\centerline{
\epsfysize=6.5cm
\epsfbox{grcfig1.eps}
}
\caption{\footnotesize
Multi peripheral diagrams.
}
\vskip 1cm
\centerline{
\epsfysize=6.5cm
\epsfbox{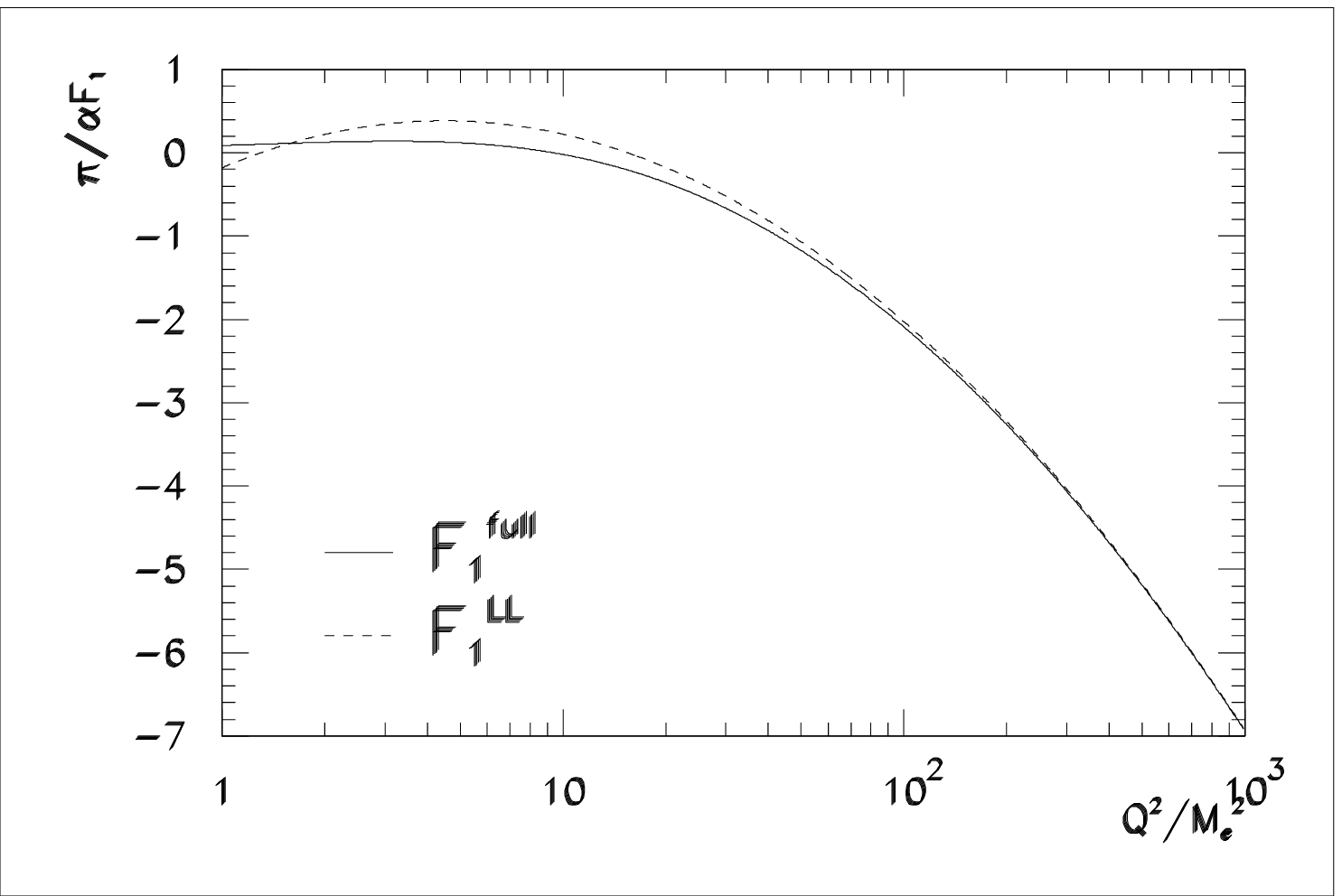}
}
\caption{\footnotesize
The exact form-factor (Re$F_2$) and that with LL approximation.
}
\end{figure}
\begin{figure}[htb]
\centerline{
\epsfysize=15cm
\epsfbox{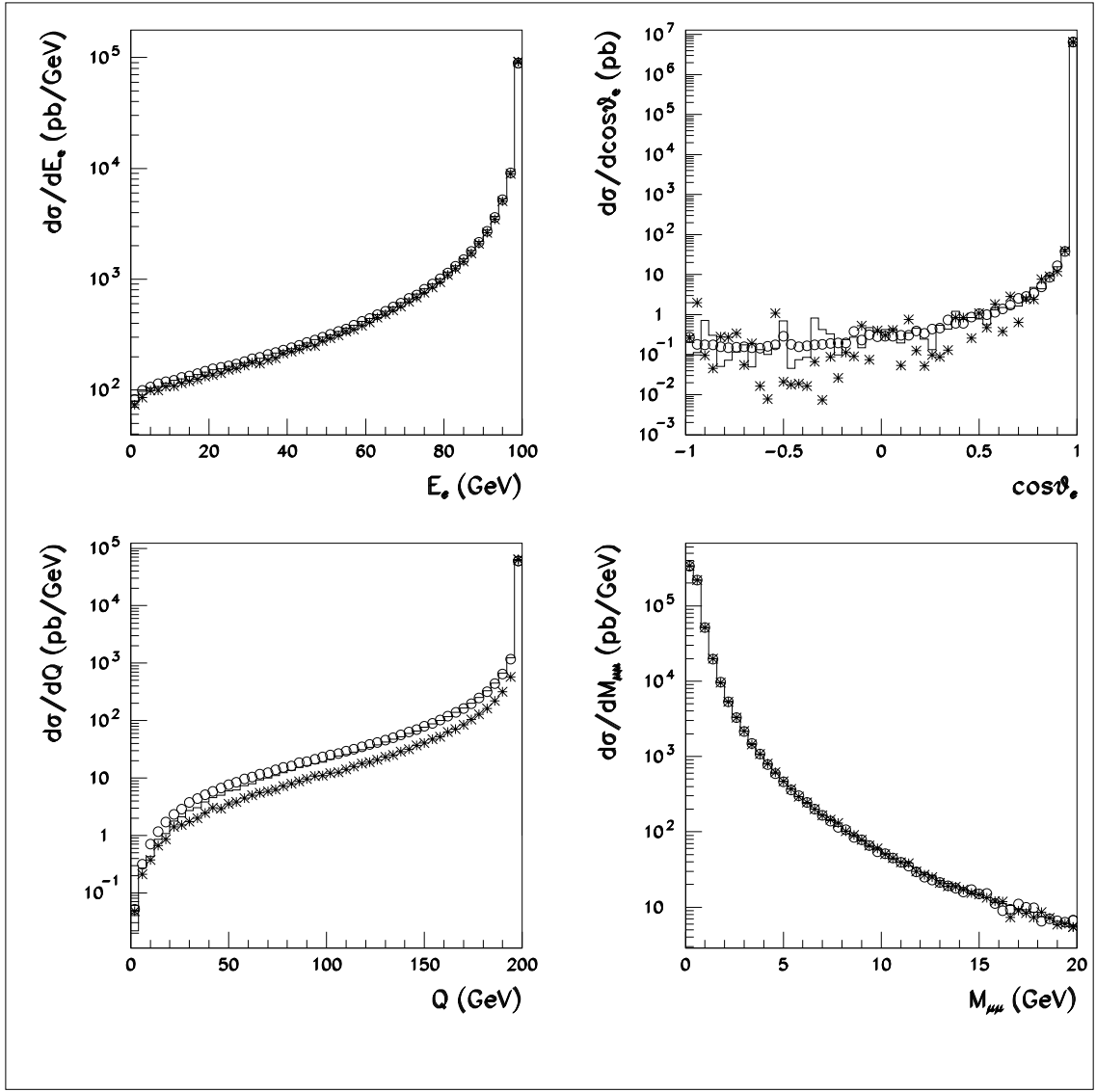}
}
\caption{\footnotesize
The differential cross sections without any experimental cuts.
The solid histograms show the {\tt GRACE-with-QEDPS} results, 
the stars the {\tt GRACE-with-SF}, and
the circles the {\tt BDK} results.
}
\end{figure}
\begin{figure}[htb]
\centerline{
\epsfysize=12cm
\epsfbox{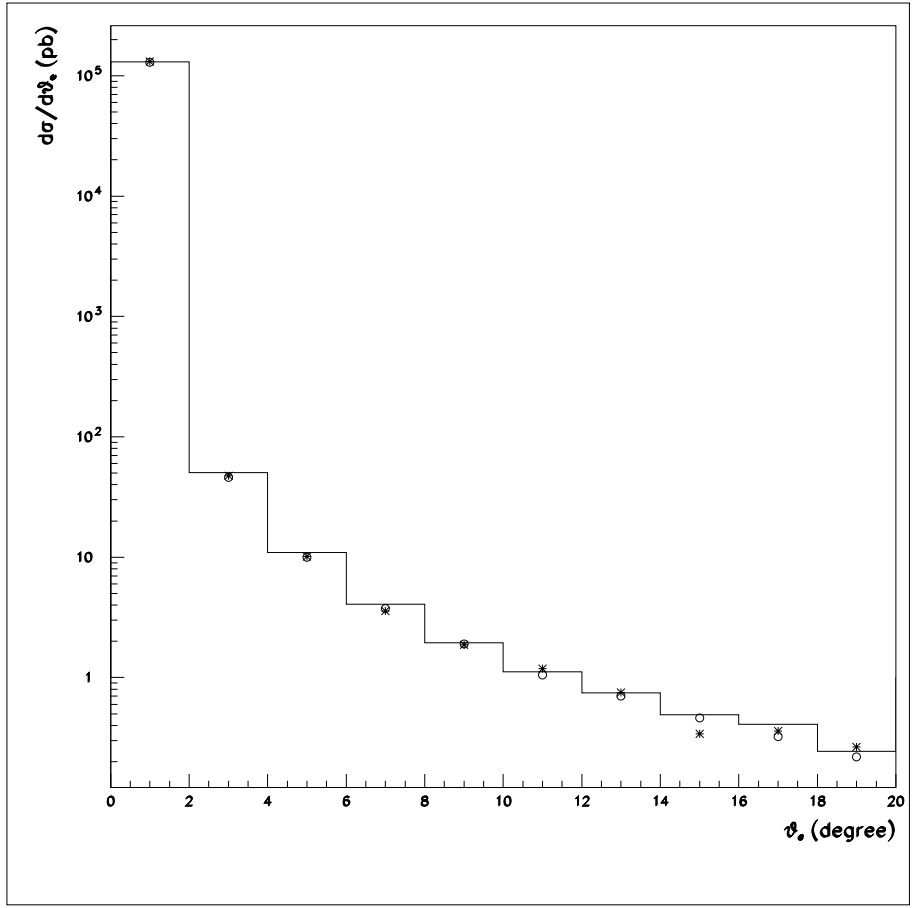}
}
\caption{\footnotesize
The $e^-$ angle distribution with respect to the beam axis 
without any experimental cuts.
The solid histograms show the {\tt GRACE-with-QEDPS} results, 
the stars the {\tt GRACE-with-SF}, and
the circles the {\tt BDK} results.
}
\end{figure}
\begin{figure}[htb]
\centerline{
\epsfysize=15cm
\epsfbox{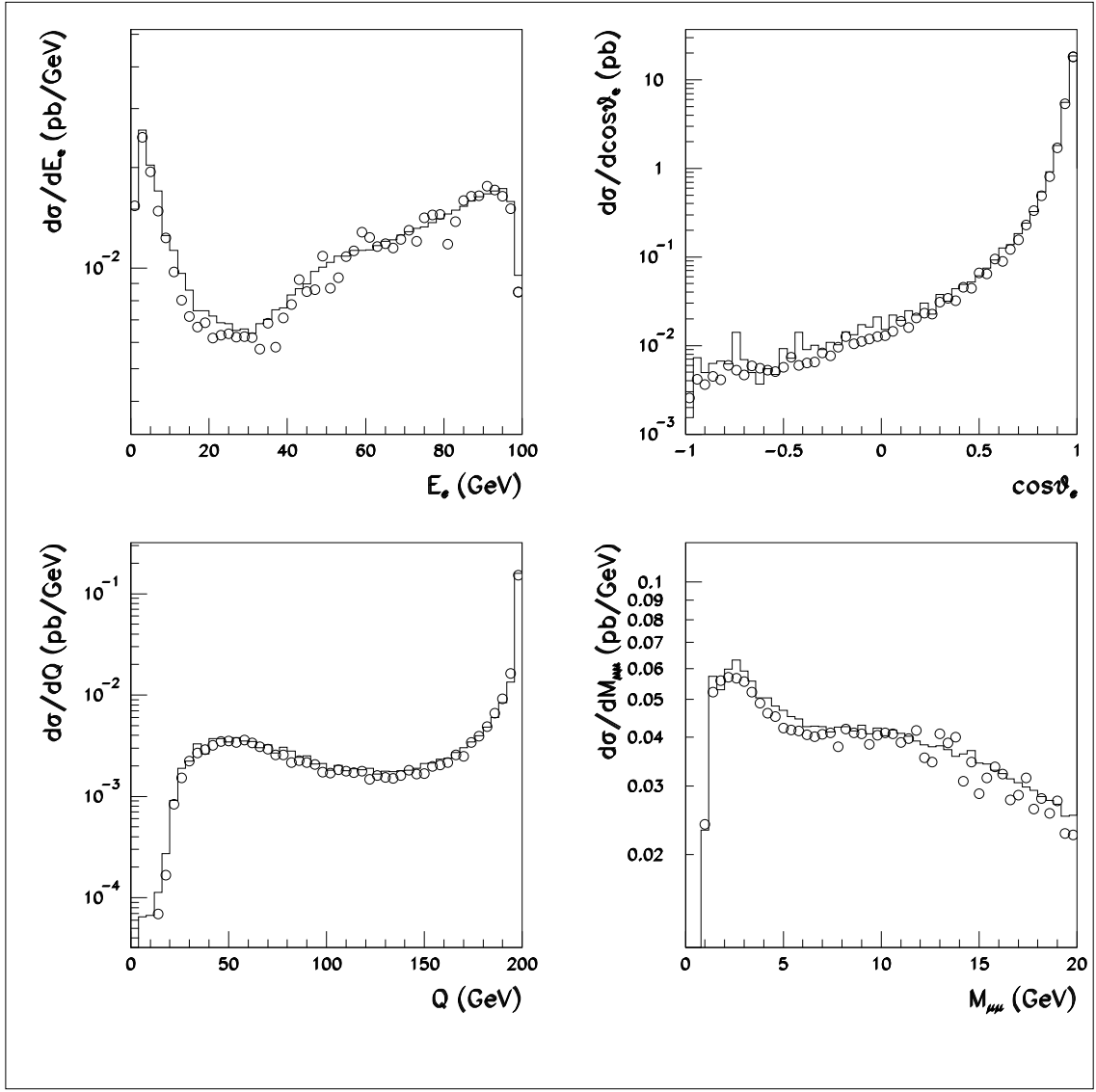}
}
\caption{\footnotesize
The differential cross sections with the $e^-$-tagging and
visible muon-pair.
The solid histograms (circles)  show {\tt GRACE} with {\tt QEDPS} 
({\tt BDK}) results.
}
\end{figure}
\end{document}